\documentclass[preprint,aps]{revtex4}
\begin{document}
\title{Minimal Modification To The Tri-bimaximal Neutrino Mixing}
\author{Xiao-Gang He$^1$ and A. Zee$^2$}
\address{$^1$Center for Theoretical Sciences, Department of Physics, National Taiwan University, Taipei, Taiwan\\
and \\$^2$Kavli Institute for Theoretical Physics, UCSB, Santa
Barbara, CA 93106, USA}

\begin{abstract}
Current experimental data on neutrino oscillations are consistent
with the tri-bimaximal mixing. If future experimental data will
determine a non-zero $V_{e3}$ and/or find CP violations in
neutrino oscillations, there is the need to modify the mixing
pattern. We find that a simple neutrino mass matrix, resulting
from $A_4$ family symmetry breaking with residual $Z_3$ and $Z_2$
discrete symmetries respectively for the Higgs sectors generating
the charged lepton and neutrino mass matrices, can satisfy the
required modifications. The neutrino mass matrix is minimally
modified with just one additional complex parameter compared with
the one producing the tri-bimaximal mixing. In this case, the CP
violating Jarlskog factor $J$ has a simple form
($|J|=|V_{e1}V_{e3}|/2\sqrt{3}$ for real neutrino mass matrix),
and also $V_{\mu i} = 1/\sqrt{3}$. We also discuss how this mixing
matrix can be tested experimentally.
\end{abstract}

\maketitle
\bigskip

\noindent {\bf {Introduction}}

The current neutrino mixing matrix from various experimental
data\cite{data,pdg} can be described by three neutrino
mixing\cite{fit1,fit}. The mixing matrix $V$ can be parameterized,
using the Particle Data Group convention\cite{pdg}, by three
mixing angles $\theta_{12}$, $\theta_{23}$ and $\theta_{13}$, and
one intrinsic CP violating phase $\delta$ for Dirac neutrinos. For
Majorana neutrinos there are additional two independent Majorana
CP violating phases. Present constraints on the mixing angles, at
the 99\% confidence level, are as the following\cite{fit}
\begin{eqnarray}
30^\circ <\theta_{12} < 38^\circ,\;\;\;36^\circ < \theta_{23} <
54^\circ,\;\;\theta_{13} < 10^\circ.
\end{eqnarray}
At present there is no experimental information about CP violating
phases.

The above data can be well fitted by the tri-bimaximal mixing of
the form
\begin{equation}
V_{tri-bi}=\left(
\begin{array}{rrr}
-{\frac{2}{\sqrt{6}}} & {\frac{1}{\sqrt{3}}} & 0 \\
{\frac{1}{\sqrt{6}}} & {\frac{1}{\sqrt{3}}} & {\frac{1}{\sqrt{2}}} \\
{\frac{1}{\sqrt{6}}} & {\frac{1}{\sqrt{3}}} & {-\frac{1}{\sqrt{2}}}
\end{array}
\right).  \label{theV}
\end{equation}
With a suitable normalization of the signs for the matrix
elements, the above tri-bimaximal mixing has $\theta_{12} =
\sin^{-1}(1/\sqrt{3}) = 35.2^\circ$, $\theta_{23} = 45^\circ$ and
$\theta_{13} = 0$. Here we have omitted a possible diagonal
Majorana phase matrix $P = Diag(e^{i\alpha_1}, e^{i\alpha_2},
e^{i\alpha_3})$ on the right. Since an overall phase does not play
a role in any physical process, only two of the $\alpha_{1,2,3}$
are physically independent.

The tri-bimaximal form for the mixing matrix was first proposed by
Harrison, Perkins and Scott\cite{hps} and further studied by
authors in Ref.\cite{xing}. Later, we independently arrived at the
same Ansatz\cite{hz1}. Many theoretical efforts have been made to
produce such a mixing pattern\cite{tribim,A41,A4z,A4hv}. Among
them theories based on $A_4$ symmetry provide some interesting
examples\cite{A41,A4z,A4hv}.

If future experimental data will find a non-zero value for
$V_{e3}$, it is necessary to modify the mixing pattern. Another
class of experimental data which may also lead to the requirement
of modifying the tri-bimaximal mixing is the observation of CP
violation in neutrino oscillations. CP violation in neutrino
oscillations is proportional to the CP violating Jarlskog
factor\cite{jarlskog} $J = Im[V_{e1}V^*_{e2}V^*_{\mu1}V_{\mu2}]$.
A non-zero $J$ is related to the non-removable phase in $V$
(``intrinsic'' CP violation). This is different than the source of
CP violation due to Majorana phases which do not show up in
neutrino oscillations. The tri-bimaximal mixing leads to $J=0$ and
therefore has no intrinsic CP violation. In this note we analyze a
simple mixing matrix\cite{A4z,A4hv}, resulted from theories based
on $A_4$ family symmetry breaking, satisfying the required
modifications.

In the modified mixing matrix, $V_{\mu i} = 1/\sqrt{3}$ which are
the same as those in the tri-bimaximal mixing. However, the matrix
element $V_{e3}$ is no longer zero, and is given by $V_{e3} = i (c
e^{i\rho} - s)$. Here $c = \cos\theta$, $s=\sin\theta$ with
$\theta$ being a new mixing angle. The phase $\rho$ is related to
phases in the neutrino mass matrix. The detailed meaning will be
given later. The modified mixing matrix in various limits reduces
to some of the forms considered in Refs.\cite{xing2,bj,lee,xing1}.
In the case $c=s=1/\sqrt{2}$ and $\rho =0$, the mixing matrix
reduces to the tri-bimaximal form. This mixing matrix has
intrinsic CP violation with the Jarlskog factor $J$ given by
$-(c^2-s^2)/6\sqrt{3}$. The two parameters $\theta$ and $\rho$ can
be determined by measuring $|V_{e3}|$, $|V_{\mu3}|$ and $J$.
Therefore this mixing pattern can be tested experimentally in
details.

The key point in obtaining the tri-bimaximal mixing pattern in
theories based on $A_4$ family symmetry is to first get the
matrices $U_l$ and $U_\nu$, which diagonalze  the charged lepton
mass matrix $M_l$ and neutrino mass matrix $M_\nu$, $U^\dagger_l
M_l U_r = D_l $ and $U^T_\nu M_\nu U_\nu = D_\nu$ (assuming
Majorana neutrinos), to have the following
forms\cite{A41,A4z,A4hv}
\begin{eqnarray}
U^\dagger_l={1\over \sqrt{3}} \left(
\begin{array}{lll}
1 & 1 & 1 \\
\omega & 1 & \omega ^{2} \\
\omega ^{2} & 1 & \omega
\end{array}
\right),\;\;\;\;
U_\nu ={1\over \sqrt{2}} \left(
\begin{array}{lll}
1 & 0 & -1 \\
0 & \sqrt{2} & 0 \\
1 & 0 & 1
\end{array}
\right),\label{magic}
\end{eqnarray}
where $\omega^3 =1$ and $1+\omega +\omega^2 =0$. Then using the
definition for the mixing matrix $V=U^\dagger_l U_\nu$ to obtain
\begin{eqnarray}
V = V_{tri-bi} V_\phi,
\end{eqnarray}
where $V_\phi = Diag(-1,1,-i)$.

For $U_\nu$, we recognize that it is just a rotation through
45$^{o}$ in the $(1-3)$ plane. Recalling that $U_{\nu }$ is
determined by requiring $U_{\nu }^{T}M_{\nu }U_{\nu }=D_{\nu }$ be
diagonal, the form of $M_{\nu }$ is
\begin{equation}
M_{\nu }=\left(
\begin{array}{lll}
\alpha & 0 & \beta \\
0 & \gamma & 0 \\
\beta & 0 & \alpha
\end{array}
\right).  \label{mnuform}
\end{equation}

\noindent {\bf Tri-bimaximal mass matrix and modifications}

Let us now briefly discuss how tri-bimaximal mixing can arise and
how it is minimally modified in $A_4$ models following
Ref.\cite{A4z}. The basic issue is that $A_4$ symmetry is broken
down to two different discrete subgroups upon charged leptons and
upon the neutrinos acquiring masses, namely $Z_3$ and $Z_2$
respectively. The clash between these two different subgroups was
called the ``sequestering problem''\cite{A4hv}. To explain the
clash, let us be more specific with Higgs mechanism supplying the
lepton masses. We emphasize, however, that the results of this
paper are not dependent on any specific model.

The two forms for $U_l$ and $U_\nu$ in eq.(\ref{magic}) are very
different. In $A_4$ theories, this requires at least two separate
Higgs sectors. We consider a case with three Higgs
fields\cite{A4z,A4hv}, $\Phi$, $\phi$ (standard model doublet) and
$\chi$ (standard model singlet). Under the $A_4$, $\Phi$ and
$\chi$ both transform as 3, and $\phi$ as 1. The standard
left-handed leptons $l_L$, right-handed charged leptons $(l^1_R,
l^2_R, l^3_R)$, and right-handed neutrinos $\nu_R$ transform as 3,
$(1'',1,1')$ and 3, respectively. We refer the readers for more
details on $A_4$ group properties to Refs.\cite{A41,A4z,A4hv,A4}.
The Lagrangian responsible for the lepton mass matrix is
\begin{eqnarray}
L &=& \lambda_e \bar l_L \tilde \Phi l^1_R  + \lambda_\mu \bar l_L
\tilde \Phi l^2_R + \lambda_\tau \bar l_L \tilde \Phi
l^3_R + H.C.\nonumber\\
&+& \lambda_\nu \bar l_L \nu_R \phi + m \bar \nu_R \nu^C_R +
\lambda_\chi \bar \nu_R \nu^C_R \chi.
\end{eqnarray}

If the vev structure is of the form $<\Phi_{1,2,3}> = v_\Phi$,
$<\chi_{1,3}>=0$, $<\chi_2> = v_\chi$, and $<\phi> = v_\phi$, one
would obtain the charged lepton mass term as
\begin{eqnarray}
&&\left (\begin{array}{lll} \bar l^1_L& \bar l^2_L& \bar l^3_L
\end{array} \right )\left ( \begin{array}{lll}
1&\omega^2&\omega\\
1&1&1\\
1&\omega&\omega^2
\end{array}\right )\left ( \begin{array}{lll}
\lambda_e v_\Phi&0&0\\0&\lambda_\mu v_\Phi&0\\0&0&\lambda_\tau
v_\Phi\end{array}\right )\left (
\begin{array}{l}l^1_R\\l^2_R\\l^3_R\end{array}\right ).
\end{eqnarray}
From the above, we can identify the charged lepton mass to be
$\sqrt{3}\lambda_i v_\Phi$, and  $U_l$ to have the ``magic'' form
in eq.(\ref{magic}). $U_r$ is a unit matrix.

The neutrino mass matrix has the see-saw form with
\begin{eqnarray}
M &=& \left ( \begin{array}{ll} 0&M_D \\M_D^T
&M_R\end{array}\right ),\;\; M_R =\left (
\begin{array}{lll}m&0&m_\chi\\0&m&0\\m_\chi&0&m\end{array}\right
),
\end{eqnarray}
where $M_D = Diag(1,1,1)\lambda_\nu v_\phi$, and $m_\chi =
\lambda_\chi v_\chi$. From this one obtains the light neutrino
mass matrix $M_\nu$ of the form given in eq.(\ref{mnuform}). One
therefore has a model for the tri-bimaximal mixing.

The vev structure of the Higgs fields breaks $A_4$, but left some
residual symmetries. The Higgs doublet $\Phi_i$ with equal vacuum
expectation values breaks $A_4$ down to a $Z_3$ generated by
$\{I,c,a\}$, and the vev of only the $\chi_2$ component to be
non-zero in $\chi$ breaks $A_4$ down to a $Z_2$ generated by
$\{1,r_2\}$. Here $a,\;c,\;r_2$ are $A_4$ group elements defined
in Ref.\cite{A4z}. We note that the charged lepton mass matrix and
the neutrino mass matrix are related to two separate Higgs
sectors, $\Phi$, and, $\chi$ and $\phi$, respectively. If there is
no communication between the two Higgs sectors, the residual $Z_3$
and $Z_2$ symmetries will be maintained. In general $\Phi$ and
$\chi$ mix in the Higgs potential, it is not possible to keep the
vev structure for $\Phi$ and $\chi$ discussed
above\cite{A4z,A4hv}. One needs to separate them from
communicating in the Higgs potential and therefore the
sequestering problem. This sequestering problem will complicate
the situation. However, models realizing such separation have been
constructed with additional symmetries\cite{A4hv}. For our purpose
here, we will assume that the sequestering problem is solved and
study the consequences.

As long as the $Z_3$ symmetry is not broken, i.e. equal vev for
$\Phi_i$, the form of $U_l$ obtained in the above is stable
against higher order corrections. Also if the $Z_2$ symmetry is
not broken, the ``12'', ``21'', ``23'' and ``32'' entries in $M_D$
and $M_R$ and therefore $M_\nu$ are prevented from getting
non-zero values. However it does not protect the ``11'' and ``22''
entries be equal\cite{A4z,A4hv}. Therefore after symmetry
breaking
%, including the possible symmetries which solving the
%sequestering problem,
a more general form of the light neutrino
mass matrix $M_\nu$ will emerge with
\begin{equation}
M_{\nu }=\left(
\begin{array}{lll}
\alpha -\varepsilon  & 0 & \beta  \\
0 & \gamma  & 0 \\
\beta  & 0 & \alpha +\varepsilon
\end{array}
\right), \label{new}
\end{equation}
rather than the $M_{\nu }$ in (\ref{mnuform}). The above neutrino
mass matrix has been obtained previously in Refs.\cite{A4z,A4hv}.

The above form of neutrino mass matrix is a minimal modification
to the one which generates the tri-bimaximal mixing in the sense
that there is just one additional complex parameter $\varepsilon$
introduced in the mass matrix. With the new $M_\nu$ the most
general form for $U_\nu$ is given by
\begin{equation}
U_{\nu }={V_\phi'}\left(
\begin{array}{lll}
1 & 0 & 0 \\
0 & 1 & 0 \\
0 & 0 & e^{i\delta}
\end{array}
\right)\left(
\begin{array}{lll}
c & 0 & -s \\
0 & 1 & 0 \\
s & 0 & c
\end{array}
\right)V_{\phi''},
\end{equation}
where $c = \cos\theta$, $s=\sin\theta$, and
\begin{eqnarray}
&&\tan^22\theta = {4|\beta|^2\over (|\alpha-\varepsilon| -
|\alpha+\varepsilon|)^2}\left (1 - {4|\alpha-\varepsilon||\alpha+
\varepsilon|\over (|\alpha-\varepsilon| +
|\alpha+\varepsilon|)^2}\sin^2\sigma\right ),\nonumber\\
&&\tan\delta = -{|\alpha -\varepsilon|-|\alpha+\varepsilon|\over
|\alpha-\varepsilon|+ |\alpha+\varepsilon|}\tan\sigma, \nonumber\\
&&\sigma = Arg(\beta) - {1\over 2}(\delta_{11} +
\delta_{33}),\nonumber\\
&&V_{\phi'}= Diag(e^{-i\delta_{11}/2}, 1,
e^{-i\delta_{33}/2}),\nonumber\\
&& \delta_{11} = Arg(\alpha-\varepsilon),\nonumber \delta_{33} =
Arg(\alpha+\varepsilon),\end{eqnarray} and
$(V^{\dagger}_{\phi''})^2$ is related to the neutrino Majorana
phases which are functions of $\alpha$, $\beta$, $\gamma$ and
$\varepsilon$.

Finally after reorganizing the Majorana phases, in the basis when
taking the limit that $\varepsilon$ goes to zero $V$ reduces to
$V_{tri-bi}$, we obtain a neutrino mixing matrix in the following
form
\begin{eqnarray}
V &=&{1\over \sqrt{3}}\left(
\begin{array}{lll}
1 & 1 &
1 \\
\omega& 1 &
\omega^2 \\
\omega^2 & 1 & \omega
\end{array} \right )
\left ( \begin{array}{lll} 1&0&0\\0&1&0\\0&0&e^{i\rho}
\end{array}\right )
\left ( \begin{array}{lll} c&0&-s\\0&1&0\\s&0&c\end{array}\right )
\left (\begin{array}{lll} -1&0&0\\0&1&0\\0&0&i\end{array}\right
)\\
&=&{1\over \sqrt{3}}\left(
\begin{array}{lll}
-(c + se^{i\rho}) & 1 &
i(ce^{i\rho}-s) \\
-(\omega c +\omega^2 s e^{i\rho})& 1 &
i(\omega^2 c e^{i\rho} - \omega s) \\
-(\omega^2 c +\omega  s e^{i\rho}) & 1 & i(\omega
ce^{i\rho}-\omega^2 s)
\end{array}
\right)=V_{tri-bi}\left (\begin{array}{lll} \cos\tau&0&i\sin\tau
e^{i\eta}\\0&1&0\\i\sin\tau e^{-i\eta}&0&\cos\tau\end{array}\right
)V_{p}\;, \label{dtribi}\nonumber
\end{eqnarray}
where $\rho = \delta -(\delta_{33}-\delta_{11})/2$.
$V_p=Diag(e^{i(\xi+\rho/2)},1,e^{i(-\xi +\rho/2)})$ with $\tan\xi
= (s-c) \tan(\rho/2)/(s+c)$, $\tan\eta = 2
\tan(2\theta)\tan(\rho/2)/(1+\tan^2(\rho/2)$, and $\sin^2\tau =
(1-2sc \cos \rho)/2$.
 In this basis,
the neutrino masses $m_{1,2,3}$ have  Majorana phases
$-2\alpha_{1,2,3}$ with
\begin{eqnarray}
&&\alpha_1 = -[Arg(c^2 |\alpha -\varepsilon| + 2 s c
|\beta|e^{i(\delta +\sigma)}+ s^2
|\alpha+\varepsilon|e^{i2\delta}) +
\delta_{11}]/2+\pi,\nonumber\\
&&\alpha_2 = -Arg(\gamma)/2,\nonumber\\
&&\alpha_3 =-[Arg(s^2 |\alpha -\varepsilon| - 2 s c
|\beta|e^{i(\delta +\sigma)}+ c^2
|\alpha+\varepsilon|e^{i2\delta}) +\delta_{11}]/2-\pi/2.
\end{eqnarray}

 The masses are given by \begin{eqnarray}
|m_1|^2 &=& \left |c^2 |\alpha -\varepsilon| + 2 s c
|\beta|e^{i(\delta +\sigma)}+ s^2
|\alpha+\varepsilon|e^{i2\delta}\right |^2\nonumber\\
&=&\left |{1\over 2} |\alpha-\varepsilon|(1+{1\over
\cos(2\theta)}) +{1\over 2} |\alpha+\varepsilon|(1-{1\over
\cos(2\theta)})e^{i2\delta}\right|^2\nonumber\\
 &=&
{1\over 2}[|\alpha-\varepsilon|^2 + 2 |\beta|^2 +
|\alpha+\varepsilon|^2
+ {1\over \cos(2\theta)}(|\alpha-\varepsilon|^2-|\alpha+\varepsilon|^2)],\nonumber\\
|m_2|^2 &=& |\gamma|^2,\nonumber\\
|m_3|^2& =& \left |s^2 |\alpha -\varepsilon| - 2 s c
|\beta|e^{i(\delta
+\sigma)}+ c^2 |\alpha+\varepsilon|e^{i2\delta}\right |^2.\nonumber\\
&=&\left |{1\over 2} |\alpha-\varepsilon|(1-{1\over
\cos(2\theta)}) +{1\over 2} |\alpha+\varepsilon|(1+{1\over
\cos(2\theta)})e^{i2\delta}\right|^2\nonumber\\
&=& {1\over 2}[|\alpha-\varepsilon|^2 + 2 |\beta|^2 +
|\alpha+\varepsilon|^2 - {1\over
\cos(2\theta)}(|\alpha-\varepsilon|^2-|\alpha+\varepsilon|^2)].
\label{masses}
\end{eqnarray}

\noindent {\bf Properties of the modified mixing matrix}

One clearly sees that the new mixing matrix can be very different
from the tri-bimaximal, but the entries $V_{\mu i}=1/\sqrt{3}$ are
the same as those of the tri-bimaximal mixing. This can be tested
in the near future by experiments. There are of course many new
features. Two important qualitative differences are:

(a). $V_{e3}$ is not zero any more. We have $|V_{e3}| = |(c
e^{i\rho} - s)|/\sqrt{3}$. In the real neutrino mass matrix case,
for small $\varepsilon$\cite{A4z},
\begin{eqnarray}
|V_{e3}| \approx {|\varepsilon|\over \sqrt{6} |\beta|}.
\end{eqnarray}

(b). There are intrinsic CP violation. This can be easily checked
by evaluating the Jarlskog factor, we obtain\cite{A4hv}
\begin{eqnarray}
J = -{1\over 9} (c^2-s^2) \sin{2\pi\over 3}.
\end{eqnarray}

It is surprising to note that  the CP violating Jarlskog factor
$J$ does not contain the phase $\rho$ implying that even if the
parameters $\alpha$, $\beta$ and $\varepsilon$ are real (or
$\rho=0$) there is intrinsic CP violation. In this case the value
of $J$ is equal to $-iV_{e1} V_{e3}/2\sqrt{3}$ whose size can be
as large as 0.04. This is sizeable enough to be measured in future
experiments.

Note that the mixing matrix, apart from the Majorana phases
$\alpha_i$, is a two-parameter, $\rho$ and $\theta$, matrix. They
can be completely determined experimentally.

The sign of $J= -(c^2-s^2)/6\sqrt{3}$ will decide whether $c$ is
larger or smaller than $s$.  We have
\begin{eqnarray} \cos 2\theta
=-6\sqrt{3}J.
\end{eqnarray}
One can always choose a convention with both $s$ and $c$ be
positive. We then have $\sin2\theta = \sqrt{1-(\cos2\theta)^2}$.

The phase factor $\rho$ can be determined by additional
measurements of $V_{e3}$ and $V_{\mu 3}$.  We have
\begin{eqnarray}
\cos\rho = {1-3 |V_{e3}|^2\over \sqrt{1-(6\sqrt{3}J)^2}}.
\end{eqnarray}
Combining \begin{eqnarray} \tan\rho = - {2\over \sqrt{3}}\left
({1\over 2}+{1-3|V_{\mu 3}|^2\over 1-3|V_{e3}|^2}\right ),
\end{eqnarray}
the sign of $\sin\rho$ can also be determined. The consistency of
the above two equations can provide tests for the mixing matrix
proposed.

We comment that the mixing matrix contains some of the cases
studied by Xing\cite{xing2}, Bjorken, Harrison and Scott\cite{bj},
Friedberg and Lee\cite{lee}, and  Xing, Zhang and Zhou\cite{xing1}
in various limiting cases. We find the following two limiting
cases interesting.

(1). $c=s=1/\sqrt{2}$. In this case there is no intrinsic CP
violation (no CP violation can be observed in neutrino
oscillations). We have from eq.(\ref{dtribi}),
\begin{eqnarray}
V = V_{tri-bi} \left (\begin{array}{lll} \cos(\rho/2)
&0&\sin(\rho/2)\\0&1&0\\-\sin(\rho/2)&0&\cos(\rho/2)
\end{array}\right )\left (\begin{array}{lll} e^{i\rho/2}&0&0\\0&1&0\\0&0&e^{i\rho/2}
\end{array}\right ).
\end{eqnarray}
This is the same mixing matrix, up to some Majorana phases, in eq.
(1.18) obtained in Ref.\cite{lee}. From this we also see that the
phase $\rho$ indeed does not play the role of a Dirac phase which
causes  intrinsic CP violation.

(2). $\rho=0$, in this case there is intrinsic CP violation. We
have
\begin{eqnarray}
V = V_{tri-bi} \left (\begin{array}{lll} \cos(\theta-\pi/4)
&0&i\sin(\theta-\pi/4)\\0&1&0\\i\sin(\theta-\pi/4)&0&\cos(\theta-\pi/4)
\end{array}\right ),
\end{eqnarray}
and $J = -iV_{e1}V_{e3}/2\sqrt{3}$. This mixing matrix looks
similar to that in eq. (1.18) of Ref.\cite{lee}, but the
appearance of ``$i$'' makes it CP violating.

It would be interesting to see how these two limiting cases can be
experimentally distinguished. An obvious way to tell the
difference is to determine whether $J$ is zero or not by measuring
CP violation in neutrino oscillations. If $J$ turns out to be
non-zero, case (1) has to be abandoned. Before $J$ can be
measured, precise measurements of $|V_{\mu 3}|$ and $|V_{e3}|$ can
also tell the difference since for case (1), one has
\begin{eqnarray}
|V_{\mu 3}|^2 = {1\over 2} (1-|V_{e3}|^2) \pm {1\over
\sqrt{2}}|V_{e3}|\sqrt{1-3|V_{e3}|^2/2},
\end{eqnarray}
where  ``$+$'' should be taken if $\cos(\rho/2)$ and
$\sin(\rho/2)$ have the same sign. Otherwise ``$-$'' should be
taken. While for case (2), one has \begin{eqnarray} |V_{\mu 3}|^2
= {1\over 2} (1-|V_{e3}|^2).
\end{eqnarray}
Since $|V_{e3}|$ is small, $|V_{\mu 3}|$ in case (2) has a weaker
dependence on $|V_{e3}|$ compared with case (1).

Measurements on quantities related to neutrino masses can also
determine the parameters in the neutrino mass matrix. We list a
few of them in the below.

i) Neutrinoless double beta decay measurement determines the
11-element $m_{ee}$ of the neutrino mass matrix in the basis where
the charged lepton mass matrix has been diagonalized. In this
model $|m_{ee}|=|2\alpha+ 2\beta +\gamma|/3$.

ii) Tritium beta effective electron neutrino mass $m_{\nu_e} =
\sqrt{\sum_i|V_{ei}m_i|^2}$ determines
\begin{eqnarray}
m^2_{\nu_e}&=&|\alpha-\varepsilon|^2 + 2|\beta|^2
+|\varepsilon+\alpha|^2+|\gamma|^2 + 4 Re(\beta \alpha^*).
\end{eqnarray}

iii) Measurements of $\Delta m^2_{12}$, $\Delta m^2_{23}$ from
neutrino oscillation data and $m_{sum} = |m_1|+|m_2|+|m_3|$ from
cosmology data can also provide information about the mass matrix
parameters using eq.(\ref{masses}). The modified mass matrix
allows both normal and inverted neutrino mass hierarchies. If
$\cos(2\theta)$ and $(|\alpha-\varepsilon|^2 -
|\alpha+\varepsilon|^2)/\cos(2\theta)
>0$ the mass hierarchy is inverted and otherwise the mass
hierarchy is a normal one.

We finally comment that to rule out the mixing proposed here it is
necessary to have precise measurement of $|V_{e2}|$. If future
experiments will determine a $|V_{e2}|$ significantly deviate from
$1/\sqrt{3}$, one has to further modify the model. In the $A_4$
based model discussed earlier, this implies that a further break
down of the residual $Z_3$ and $Z_2$ must happen. If just $Z_3$ is
broken, the vev of the components $<\Phi_i>$ will not be equal,
this will affect the ``magic'' form $U_l^\dagger$ in
eq.(\ref{magic}), whereas if  $Z_2$ is broken, the zero entries in
eq.(\ref{new}) will become non-zero. In general both $Z_3$ and
$Z_2$ may be  broken at the same time. The form of the mixing will
become the most general one with corrections to all
elements\cite{ddd}.  We have to wait more precise data to tell us
if the simple mixing proposed here need to be further modified.

\noindent {\bf {\large Acknowledgments\medskip }}

AZ thanks the hospitality of the Department of Physics at the
National Taiwan University where this work was initiated. This
work was supported in part by NCTS. XGH was supported by Taiwan
National Science Council. AZ was supported in part by US National
Science Foundation under grants PHY 99-07949 and PHY00-98395.

\smallskip

{\bf {\large References}}

\end{document}